# EXTRACTION OF TECHNICAL INFORMATION FROM NORMATIVE DOCUMENTS
# USING AUTOMATED METHODS BASED ON ONTOLOGIES :
## Application to the ISO 15531 MANDATE standard - Methodology and first results

A.F. Cutting-Decelle(*), A. Digeon(***), R.I. Young(****), J.L. Barraud(**), P. Lamboley(**)

(*) University of Geneva, CH / Ecole Centrale de Lille, F
(**) Schneider Electric, F
(***) Previous ISO TC 184 Chairman
(****) University of Loughborough, UK

**Abstract**: problems faced by international standardization bodies become more and more crucial as the number and the size of the standards they produce increase. Sometimes, also, the lack of coordination among the committees in charge of the development of standards may lead to overlaps, mistakes or incompatibilities in the documents.

The aim of this study is to present a methodology enabling an automatic extraction of the technical concepts (terms) found in normative documents, through the use of semantic tools coming from the field of language processing.

The first part of the paper provides a description of the standardization "world", its structure, its way of working and the problems faced; we then introduce the concepts of semantic annotation, information extraction and the software tools available in this domain. The next section explains the concept of ontology and its potential use in the field of standardization.

We propose here a methodology enabling the extraction of technical information from a given normative corpus, based on a semantic annotation process done according to reference ontologies. The application to the ISO 15531 MANDATE corpus provides a first use case of the methodology described in this paper.

The paper ends with the description of the first experimental results produced by this approach, and with some issues and perspectives, notably its application to other standards and / or Technical Committees and the possibility offered to create pre-defined technical dictionaries of terms.

## Introduction

The overall aim of standardization is to contribute to the simplification of users' life by providing ways of defining, specifying and structuring certain fields of knowledge. Outside the telecom world, two international bodies are in charge of the drafting of "de jure" standards: IEC in the electrical domain and ISO for the other domains.

In the industrial field, IEC and ISO are complementary. To take a simple industrial example, PLC (Programming Logic Controller) at the electrical device level as well as environmental tests, programming languages... generally comply with an IEC standard (i.e. IEC 61131, under the responsibility of IEC Technical Committee 65) but when it comes to the design, sourcing, manufacturing, ... and disposal of industrial systems or the integration of multiple industrial systems, ISO (and namely its Technical Committee ISO TC 184 "Automation Systems and Integration") comes into play.

Overall, this way of organizing fields of knowledge has proven successful. Just for ISO TC 184, over 817 standards have been published over the years, each under the direct responsibility of ISO TC 184. But, sometimes, users of standards may be confronted with consistency issues. The simple definition of a "user", or of a "resource" in a given standard may not be exactly the same as its definition in another standard, even when defined by the same Technical Committee.

For someone wanting to take advantage of standards for structuring and implementing a complex system, it may become cumbersome, as no coherent nor global vocabulary or directory of concepts exists. This is why, for many years, succeeding ISO TC 184 chairmen, with the help and contribution of Sub Committees and experts have worked to bring more consistency within the scope of their field of standardization.

Note that having users of standards as the key benefiters of this work implies, in the case of ISO TC 184, to share its views and collaborate with IEC TC 65. When this effort toward increased consistency started, it was all manual. Excel spreadsheets were used to gather experts' viewpoints, disseminate information and present results. This was better than nothing but very time consuming for the experts and not very efficient. It proved to be laborious, as no one likes this kind of work. Not an ideal approach when dealing with hundreds of standards!

How convenient would it be if, when starting work on a new standard, experts could access consistent and shared definitions of certain concepts and identify other standards closely related to this New Work Item proposal! Expert's time, a rare and very valuable resource, would then be used in a more efficient way.

Fortunately, technologies evolve and we have now reached a stage where we can envisage automating at least part of this process. ISO TC 184 is leading this effort, and is proposing a new and effective method of bringing efficiency and consistency into the field of standardization.

Such an effort would clearly benefit standard's users, as they know, from the start of their work in using certain standards, that there will be consistency errors between the standards, or areas needing their own expert's attention and therefore technical reconciliation of inconsistencies (if and when feasible).

For the experts working on defining additional standards, there would be an immediate, free and efficient access to concepts already defined, known and used by their peers in their or other Technical Committees. This would be a great contribution to minimizing interoperability issues, and a basis for addressing real user's need for standards without reinventing wheels. And obviously, up to now, the more standards, the more risks of inconsistencies. Only the user will discover such inconsistencies, and this is far too late in the value-added chain.

This paper is mainly focused on industrial systems: communications and interoperability, as dealt with by ISO TC 184. The scope of ISO TC 184 is *standardization in the field of automation systems and their integration for design, sourcing, manufacturing and delivery, support, maintenance and disposal of products and their associated services*. Areas of standardization include information systems, automation and control software and integration technologies. ISO TC 184[1] is composed of three SC's and one Advisory Group (AG). Among the standards developed by SC4, let us just mention : ISO 8000 (data quality), ISO 10303 STEP (STandard for the Exchange of Product model data), ISO 13584 P-LIB (Part Library), ISO 15531 MANDATE (MANufacturing DATa Exchange), ISO 15926 Oil and Gas, ISO 16739 IFC (Industry Foundation Classes) and ISO 18629 PSL (Process Specification Language).

This paper is structured as follows: the first part presents standardization at the international level, the standardization bodies and some of the challenges faced by standardization. Then, we give the main principles of information extraction and semantic annotation based on ontologies. We go on describing the main features of the software tools used for this study. The aim of the next section is to identify the ontologies applicable to the industrial domain, and particularly for the semantic annotation of the ISO 15531 MANDATE standard. The next part outlines the methodology developed to annotate the standard, and some of the first findings of this on-going study. Some of the benefits of the use of such semantic approaches to standards are presented, together with possible future research actions in this domain.

## 1- Standardization at the international level (ISO, IEC and ITU)

We will not address below standards emerging from user's groups or industries, even if they are valuable, but we will focus on the way standards targeting an international approval get generated and approved, as these standards are more and more required within the more and more internationally oriented business environment.

They are approved or adopted by one of the **National**, **Regional** or **International** standards bodies, mainly:

- IEC (International Electrotechnical Commission): develops international standards for electricity, electronics and related technologies - the generic term for this domain is "electrotechnologies".
- ISO (International Organization for Standardization): develops standards in the fields of industry, trade, and services. ISO is a worldwide federation of national standards bodies (ISO member bodies). The preparation work of International Standards is carried out through ISO Technical Committees (TC). ISO collaborates closely with the International Electrotechnical Commission (IEC) on all matters of Electrotechnical standardization. Most of

---

the TC's in ISO are subdivided into Sub-Committees (SC), whose role is to refine the standardization activity of the corresponding TC, focusing on specific aspects of the domain covered by the TC.
- ITU (International Telecommunication Union): United Nations specialized agency for information and communication technologies – ICTs

We focus here on ISO, since it provides the cradle of most of the standards mentioned in this paper.

## 1-1 Main principles: standard, standardization (ISO)

ISO[2] is an independent, non-governmental international organization with a membership of 161 national standards bodies and 780 technical committees and subcommittees to take care of standards development. Over 135 people work full time for ISO Central Secretariat in Geneva, Switzerland. Through its members, it brings together experts to share knowledge and develop voluntary, consensus-based, market relevant International Standards that support innovation and provide solutions to global challenges. Agreements have been made with the European Committee for Standardization (CEN) to create a sort of "non-aggression" pact between the two entities and to avoid redundant standards. ISO cooperates also with the International Electrotechnical Commission (IEC), responsible for the standardization of electrical equipment. The most important landmark of this cooperation was, in the mid-1980s, the creation of the Joint Technical Committee on Information Technologies (JTC1) which brought together the technical competence held by IEC and the software competence held by ISO. JTC 1 is the producer of most of the standards for computing. ISO defines the standard as a *"Document established by consensus and approved by a recognized organization, which provides, for common and repeated uses, rules, guidelines or characteristics, for activities or their results ensuring an optimal level of order in a given context."*

- **Why we need standards:** Standards are needed to ensure product safety, fitness for the use of products and materials, to promote the interoperability of products and services, facilitating trade by removing barriers to trade and to promote a common understanding of what is a "product".
- **Who develops standards:** Standards are developed by experts from all socio-economic international sectors. They provide the technical content of normative documents and their updating within national standardization committees and international technical committees.

- **ISO members:** three categories of members, with different levels of access and influence in the ISO system:

- **Full members** influence standardization work and ISO strategies. They are empowered to participate with full voting rights in all ISO technical and political meetings. Full members sell ISO International Standards and can adopt them as national standards**.**
- **Corresponding** members observe the development of ISO standards and the strategies. They are entitled to attend technical and political meetings as observers. Corresponding members sell ISO International Standards and can adopt them as national standards**.**
- **Subscriber members** are kept informed of ISO activities but cannot participate. They are not allowed to sell ISO International Standards or adopt them as national standards**.**

- **ISO deliverables:** There are different types of ISO deliverables:

- **ISO Standard:** An International Standard provides rules, guidelines or features relating to activities or their results, with the aim of achieving the optimal degree of order in a given context, under many forms: product standards, but also test methods, codes of practice, guidelines and management systems standards.
- **ISO / TS (Technical specification):** work still in the technical development phase or estimated to be subject to future agreement on an International Standard. A Technical Specification is published for immediate use, but also to obtain feedback.
- **ISO / TR (Technical report):** contains information that differs from that of the two previous publications. It may include, for example, data from a survey or informative report, or information about the actual "state of the art".
- **ISO / PAS (Publicly Available Specification):** published to address an urgent market need and represents either the consensus of experts within a working group, or a consensus in an organization outside the ISO.
- **International Workshop Agreement (IWA):** document developed outside the ISO system, to allow market players to negotiate in an "open workshop" environment. They usually benefit from the administrative support of a committee member.
- **ISO Guide:** helps readers to better identify the main areas where standards bring added value. Some guides explain how and why ISO standards help to improve a product or process, to make it safer and more efficient.

- **Basic principle of standard setting:** ISO standards respond to a market needs: ISO does not launch by itself the

---

[2] https://www.iso.org/fr/home.html

development of a new standard. ISO responds to a demand expressed by industry or other stakeholders such as consumer associations. Typically, a sector or group signals the interest of a standard to the ISO member for their country, who informs ISO. ISO standards are based on global expertise: ISO standards are developed by groups of experts from around the world regrouped in groups: the Technical Committees. Experts negotiate standards in every detail, including the scope, key definitions, and content. ISO standards are the result from a multi-stakeholder process: the technical committees are made up of experts from the industries concerned, but also representatives of consumer associations, academia, NGOs and governments. ISO standards are based on consensus: The development of ISO standards can be seen as a synthesis of a consensual approach and the observations of stakeholders are taken into account.

It is worth noting here that, if ISO, IEC, ITU and other Standardization Bodies draft "de jure" standards, other groups, either non-profit organizations, or professional associations (ISA, IEEE, OMG, OASIS, W3C and others) also contribute to the development of standards, called "de facto" standards since they are not authored by statutory bodies.

**- Standards development process**: Any ISO member national body, any committee, any organization linked to ISO may have the initiative to modify/create a standard, according to the technological, economic and societal developments (see: Developing ISO standards[3]). At the outset, each ISO deliverable is assigned to a *standards development track*. This track determines the timeframe of the project as it passes through the various *stages* to publication. Stages and resources for standards development are: (* = *obligatory stage)* 1. Proposal stage*; 2. Committee stage*; 3. Enquiry stage*; 4. Approval stage*; 5. Publication stage*. Documents to be provided at each stage of the standards development process, and conditions for moving from one stage to another are documented in the ISO Directives, Part 1 and Part 2 (ISO/IEC Directives, Part 1, 2017), (ISO/IEC Directives, Part 2, 2016), (ISO/IEC Directives, Consolidated ISO document, 2017).

A way of benefiting from standardization is also for large industrial enterprises or professional associations to develop documents, sets of procedures, internal guidelines, following as closely as possible the structure of normative documents (all the easier when those enterprises are already involved in several standardization committees and/or working groups!), then submit their document to FastTrack procedures in order to get the status of IS (or other) as quickly as possible.

## 1-2 Technical committees and sub-committees - Chairmanship - challenges

More than 22 156 International standards in almost all fields of technology and economics have been published by ISO till now. Those standards are developed by 780 technical committees and subcommittees.
It is very important to understand why countries seek to assume the responsibilities of international technical committees or subcommittees (end 2016)[4] : Every year a standardization report is sent to DGE (Directorate-General for Enterprise). This report highlights the advantages of an active participation in the international Standardization Bodies in order to defend national economic interests and promote economic models: France, Germany, UK and USA have still a strong presence in the management of ISO technical committees and subcommittees responsibilities, but, to name just 2 other countries, Japan and China are increasing their participation, to target new challenges, especially projects in new technologies fields.
For the experts (chosen on the basis of their technical skill), the goal is to develop standardization projects and to actively contribute to ISO working groups. ISO technical committees and subcommittees Chair's must also respect certain rules, among which: the respect of objectives and deadlines, according to normative ISO rules; to ensure that for any new topic, issues and objectives have been identified; to conduct the work in order to obtain a consensual agreement; to ensure that the technical levels of the standards are consistent with those of the sector concerned and economic constraints are taken into account. The benefit is to promote their company, to work with international partners and to represent their industrial sector.

**- Establishment of technical committees[5] :** Technical committees are established and dissolved by the Technical Management Board. TMB may transform an existing subcommittee into a new technical committee, following consultation with the technical committee concerned. The technical management decides the establishment of a new technical committee, provided that a 2/3 majority of the National Bodies voting are in favor of the proposal, and at least 5 National Bodies who voted in favor express their intention to participate actively, and allocate the secretariat or assign the work to an existing technical committee, subject to the same criteria of acceptance: Fig. 1.

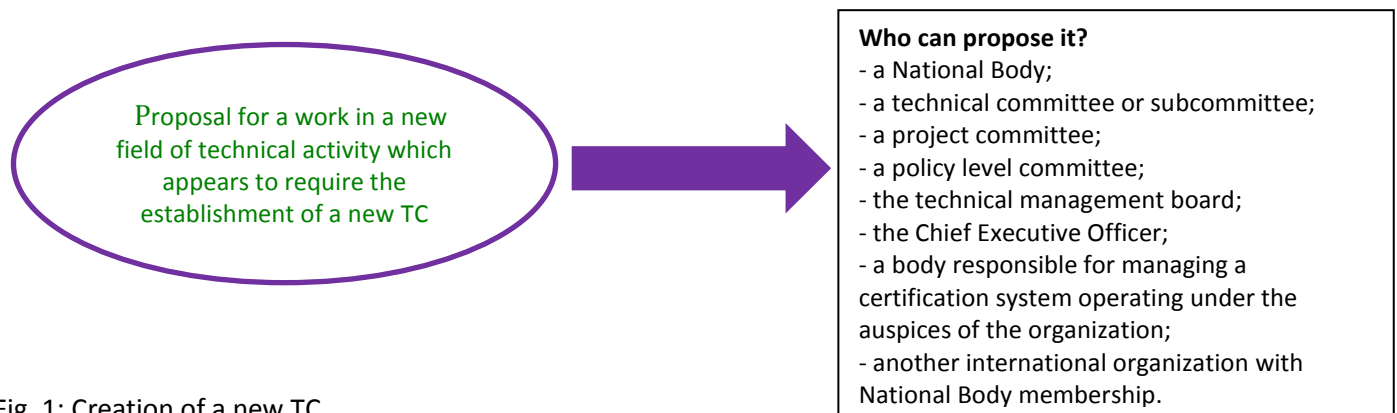

Fig. 1: Creation of a new TC

Once the new Technical committee validated and the secretariat appointed, the National Standardization Body concerned shall appoint a chairperson, based on his/her ability to manage the strategic and economic issues of the field concerned. On the other hand, a technical committee must be funded by stakeholders. Generally, the company who wishes to obtain the technical committee leadership becomes the main funder or in other cases, the funding is pooled with other actors (concerned parties)[6].

## 1-3 ISO Technical committees and subcommittees - real life and problems met

Standardization is a "big machine", with many challenges, particularly through functions and roles countries play in TC's or SC's (particularly for chairman and secretary). For a country, the fact of taking the leadership makes it possible to take its economic model to the international level, since the chairman and the secretary contribute to guide the strategy of the whole technical committee or subcommittee. When a Technical Committee is created, an international meeting is held, usually in the country holding the chairmanship and the secretariat. Many resolutions are taken: scope validation, work program, creation of working groups, appointment of Convenors, schedule of meetings... A technical committee lives if and only if its work program is active, the objective being to become a reference in its field of standardization.

However, the reality of the standards development process is not simple: ISO TC's develop standards, they propose draft standards that are generally accepted, but many issues arise at this stage. We will mention some of them here, one of the aim of this paper is to propose elements to help solving some of those issues through the methodology developed.

About the problems faced, many questions can be asked:
- Is there a search for relevance or redundancy with other on-going developments?
- Who validates the new project proposals? The standardization offices vote, accept the project, but they do not have the possibility to check the consistency, since more than 22 156 standards have been published by ISO.
- Who ensures that those projects are not developed in another TC? The chairs of TC's rely on the WG Convenors and their expertise in the field concerned, however they do not have visibility on the scope of all the TC's.
- The development of new standards requires, from the experts involved, a wide skill and knowledge of the domain in order to avoid gaps/overlaps among the standards. NSB and SDOs have recently identified a real issue as ISO is not able to verify whether the experts, who developed the projects, have taken a moment to read the previous parts again, to search in the different standards on the same topic their relevance to the new development.
- Who reads the standards to see if they are consistent? Many standards have been developed, some of them in several parts, as shown by ISO/8000 - Data quality: 350 parts to its credit. It is no easy task for experts to be able to guarantee that the standard really meets the market need at the time it is published. Or to be confident that all parts are consistent with one another. We will show in this paper some examples of mistakes we identified in the ISO 15531 MANDATE standard subject of this study.

---

[6] person or body (3.2.1) that may have an impact, be affected or have a point of view likely to affect them by a decision or activity".

- Some of the standards are very « big » in terms of number of pages, it is not easy to check the different documents against possible inconsistencies, mistakes …
- Another important point to mention: it becomes more and more difficult to fund TC's – it will become also more and more difficult to find good experts and get a share of their time!  The world has changed, everything is moving faster and faster and it is necessary to produce documents in a very short time, which does not leave much time for relevant proofreading of projects in development or published standards.

Whence the need to rely on automated approaches to provide common representations of the concepts handled in the normative documents. Ontology based information extraction can be seen as an answer to this need; this will allow to efficiently search the international normative corpus in order to develop coherent projects. This would contribute to make international standardization more "digestible" for the end-user, in perfect accordance with the first commitment of standardization: speak the same language.

In the next section, we provide a way to index/annotate specific information contained in normative documents.

## 2- Information extraction and semantic annotation: overview and main principles

We recall here the main definitions applicable to semantic annotation, as a particular aspect of Information extraction, then the information extraction based on ontologies. In terms of software tools applicable to information extraction, the study presented here is carried out with GATE, one of the most widely used software tool for text engineering. This section ends with a quick description of the ISO 24617 standard, since the Part 6 of this standard deals with semantic annotation.

### 2.1 Definitions and main principles

- **Information Extraction (IE)** is a technology based on the analysis of natural language in order to extract snippets of information (Bontcheva *et al.*, 2006). The process takes texts (and sometimes speech) as input and produces fixed format, unambiguous data as output. This data may be used directly for display to users, or may be stored in a database or spreadsheet for later analysis, or may be used for indexing purposes in information retrieval (IR) applications such as internet search engines like Google. However IE is quite different from IR:
- an IR system finds relevant texts and presents them to the user;
- an IE application analyses texts and presents only the specific information from them that the user is interested in.
Generally speaking, information extraction (IE) is about finding five different types of information in natural language text:
  - entities : things in the text, for example people, places, organisations, amounts of money, dates, often referred to as Named Entities (NE), …
  - mentions: all the places that particular entities are referred to in the text
  - descriptions of the entities present
  - relations between entities
  - events involving the entities.
These various types of IE provide progressively higher-level information about texts (Cunningham, 2005).

- **Annotation**: In the Oxford Online Dictionary[7], the term of "annotation" is defined as "*a note by way of explanation or comment added to a text or diagram*" (Liao *et al.*, 2011). It is used to enrich the target information object, through the use of textual descriptions, comments, tags, highlights, images, links…  An annotation may have different meanings and usages according to the domain of use.
Bechhofer *et al.* (2002) and Boudjlida *et al.* (2006) distinguish annotations as : (i) Textual annotation : adding notes and comments to objects ; (ii) Link annotation : linking objects to a readable content ; (iii) Semantic annotation : consisting in semantic information that is machine readable.
Similarly, three types of annotations are described by Oren, et al. (2006): (i) Informal annotation: notes that are not machine readable ; (ii) Formal annotation : notes that are formally defined and machine readable (but it does not use ontology terms) ; (iii) Ontological annotation : notes that use only formally defined ontological terms that are commonly accepted and understood.

---

[7] http://oxforddictionaries.com

Bechhofer, et al. (2002) classify the annotation according to six possible uses that are not always clear nor disjoint: (a) Decoration, comments on an object; (b) Linking, link anchors; (c) Instances identification, strong assert that an object is an instance of a particular class. It may use a URL; (d) Instance reference, less clear than instance identification, reference depending on background and world knowledge; (e) Aboutness, loose association of the object with a concept; (f) Pertinence, assertions about the concepts within an ontology without encoding that information.

**- Semantic Annotation**: according to the above classification, a semantic annotation can be considered as a kind of formal metadata, which is both machine and human readable. The term "Semantic Annotation" is described both as the action and the results of describing (a part of) an electronic resource by means of metadata whose meaning is formally specified in an ontology (electronic resource that can be text content, images, video, services…). For Talantikite, et al. (2009) "An annotation assigns to an entity, which is in the text, a link to its semantic description. A semantic annotation is referent to an ontology". For Lin (2008), a semantic annotation is concerned with "an approach to link ontologies to the original information sources".

Those definitions, although coming from different authors, share a common concept, the ontology: semantic annotation is the process by which an electronic resource is associated with (or linked to) a specific ontology. An ontology, considered as a specialized reference dictionary, is here the only possible means of providing a given document with formal semantics.

An example of semantic annotation is represented on Fig. 2 below:

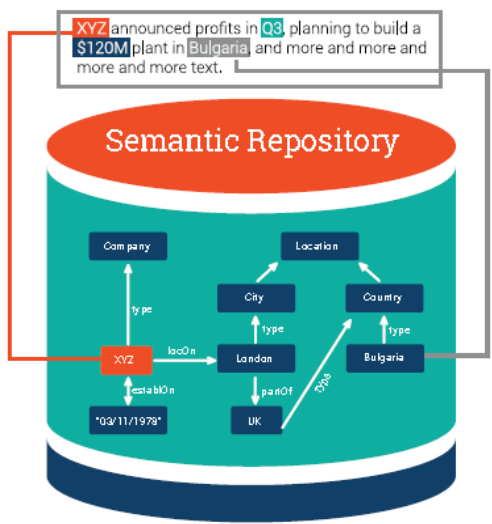

Fig. 2: Example of semantic annotation of a text (Ontotext 2016)

Terms of interest within the text are "tagged" according to given criteria, provided in a semantic repository (the ontology). This enables the enrichment of the unstructured or semi-structured data with a context that is further linked to the domain structured knowledge. It allows results that are not explicitly related to the original search (Ontotext 2016): discovery of new concepts, thus leading to a potential enrichment of the initial ontology.

In this paper, we will focus on the concept of ontology based information extraction, and its application to the technical information contained in standards.

**- Ontology:** Gruber (Gruber, 1992) defines an ontology as an explicit specification of a conceptualization. Guarino (1998), while agreeing with Gruber, presents a refined distinction between an ontology and a conceptualization: an ontology is a logical theory accounting for the intended meaning of a formal vocabulary, i.e., its ontological commitment to a particular conceptualization of the world. The intended models of a logical language using such a vocabulary are constrained by its ontological commitment. This commitment and the underlying conceptualization are reflected in the ontology by the approximation of these intended models. Guarino (1998) advises against using ontology just as a fancy name denoting the result of activities like conceptual analysis and domain modeling.

Although the use of ontologies started with Artificial Intelligence, ongoing research on ontology can be found throughout the computer science community, in such areas as computational linguistics and database theory. It covers fields ranging from knowledge engineering, information integration, and objected-oriented analysis to such applications as medicine, mechanical engineering, and GIS (Fonseca *et al.*, 2000).

Several software tools have been developed and can be used to represent ontologies. We use here the tool called Protégé v5.2[8] . Protégé Desktop is a feature rich ontology editing environment with full support for the OWL 2 Web Ontology Language, and direct in-memory connections to description logic reasoners like HermiT and Pellet. Protégé Desktop supports creation and editing of one or more ontologies in a single workspace via a completely customizable user interface. Visualization tools allow for interactive navigation of ontology relationships.

## 2.2 Ontology based information extraction

Ontology-Based IE (OBIE) is the technology used for semantic annotation (Bontcheva *et al.*, 2006). One of the important differences between traditional IE and OBIE is the use of a formal ontology as one of the system's resources. OBIE may also involve reasoning.

Another substantial difference between the semantic IE process and the traditional one is the fact that it not only finds the (most specific) type of the extracted entity, but it also identifies it, by linking it to its semantic description in the instance base. This allows entities to be traced across documents and their descriptions to be enriched through the IE process. When compared to the 'traditional' IE tasks, the first stage corresponds to the Named Entities (NE) task and the second stage corresponds to the CO (co-reference) task. Given the lower performance achievable on the CO task, semantic IE is in general a much harder task.

OBIE poses two main challenges: the identification of classes from the ontology in the text; the automatic population of ontologies with new classes in the text.

### 2.2.1. Identification of classes from the ontology
If an ontology is already populated with classes, the task of an OBIE system may be simply to identify classes from the ontology in the text. Similar methodologies can be used for this as for traditional IE systems, using an ontology rather than a flat gazetteer. For rule-based systems, this is relatively straightforward. For learning-based systems, however, this is more problematic because training data is required. Collecting such training data is, however, likely to be a large bottleneck. Unlike traditional IE systems for which training data exists in domains like news texts in plentiful form, thanks to efforts from MUC (Message Understanding Conference) (SAIC, 1998), ACE (ACE, 2004) and other collaborative and/or competitive programs, there is a dearth of material currently available for semantic web applications. New training data needs to be created manually or semi-automatically, which is a time consuming and onerous task, although systems to aid such metadata creation are currently being developed.

### 2.2.2. Automatic Ontology Population
In this task, an OBIE application identifies classes in the text belonging to concepts in a given ontology, and adds them to the ontology in the correct location. It is important to note that classes may appear in more than one location in the ontology because of the multidimensional nature of many ontologies and/or ambiguity which cannot or should not be resolved at this level (Felber, 1984, Bowker, 1995).
The transformation of texts into formal representations of the contained knowledge, in terms of annotation, paves the way for the application of sophisticated computational methods and hence helping the researchers and advance science (Gurinder Pal Singh Gosal, 2015). There have been many research efforts to build annotation systems: we will here present the main features of the software tools available.

## 2.3 Software tools

The information extraction process is twofold (Oliveira, 2013), both in terms of the method to follow for identifying the parts of the document to annotate, and in terms of the content of the annotations (Bettencourt *et al.*, 2006). We can subdivide this dimension into the type of method and the level of automation. These two sub-dimensions are related since the automation degree of the IE process is influenced by the type of method. Pattern matching methods are normally easily automated while Natural Language Processing (NLP) and ontology based methods are harder to automate.

---

[8] https://protege.stanford.edu/

The level of automation of the method ranges from manual to fully automated. Manual annotation can be achieved using diverse authoring tools that normally provide an integrated environment for authoring and annotating text. A fully manual annotation process is too expensive to carry out without a kind of automation. Depending on the technical domain, it is almost impossible to deal with the volume of available documents, on the other hand it often introduces errors, mainly due to the following factors (Bayerl et al., 2003): use of highly complex coding schemas; inconsistencies in the labeling among different annotators; and lack of familiarity with the domain.

This kind of annotation has brought the knowledge acquisition to a situation where it is difficult to facilitate the dissemination of a corpus of knowledge. To overcome this bottleneck in annotation acquisition, systems that can lead the process automatically have been proposed.

Commercially available systems propose mainly semi-automated approaches while fully automated systems are still a challenge. These systems provide the scalability needed to annotate existing documents and facilitate the annotation of new documents. These systems also facilitate the use of multiple ontologies to annotate the same document.

The type of method can be organized based on the classification presented in (Reeve, 2005):

• Pattern matching : based on regular expressions defined manually before searching the content of the document or by pattern discovery, mostly following the basic method outlined by (Brin, 1999), where an initial set of entities is defined and the corpus is exploited to discover patterns in which the entities exist. New entities are discovered along with new patterns and the process continues recursively until the method does not discover any entity or the user stops it. This method should be used on pages that do not suffer changes often.
• Rule based methods: the rules must be defined manually before searching the content of documents. This approach does not require training data and has good results on structured documents with clear patterns.
• Wrapper induction methods: defined by (Kushmerick, 1997) as the task of learning a procedure for extracting tuples from a particular information source of examples provided by the user.

Hybrid approaches are normally used to take the advantage of the strengths of methods from different categories, based on the target domain, the semantic complexity of annotations and the availability of trained human annotators and/or language engineers (Bontcheva, 2011). The real challenge of semantic annotation is to develop tools capable of a fully automated annotation.

 The concept of semantic annotation has generated a very important number of software tools, as a basic brick of the semantic web. Among those tools, some of them result from research work, or EU (or industrial) projects, or else have been developed by software companies. It is also important to note that some of those tools are dedicated to specific domains of knowledge (medicine, biology, newspapers reading, images…), others are more "neutral", agnostic with respect to a given domain. Among those software tools, we can mention: Annotator[9], KIM[10], Text2Onto[11], GATE, Proxem[12], Open Calais[13], Chimaera[14], OpenNLP[15], UIMA[16] …
The work presented in this paper is based on one of the GATE[17] software.

## 2.4 The GATE framework

GATE (Cunningham, 2002) is an open source framework for processing human language, which provides the architecture and the framework environment for developing and deploying natural language software components. Offering a rich graphical user interface, it provides easy access to language, processing and visual resources that help scientists and developers to produce natural language processing applications.
JAPE (Java Annotation Pattern Engine) is a finite state transducer, which uses regular expressions for handling pattern-matching rules. Such expressions are at the core of every rule-based IE system aimed at recognizing textual snippets that conform to particular patterns while the rules enable a cascading mechanism of matching conditions that is usually referred as the IE pipeline. JAPE grammars are constituted from two parts; the LHS (Left Hand Side) which handles the regular expressions and the RHS (Right Hand Side) which manipulates the results of the matching conditions and defines the semantic annotation outcome.

---

[9] http://annotatorjs.org/
[10] http://www.ontotext.com/kim
[11] https://code.google.com/p/text2onto/
[12] https://www.proxem.com/
[13] http://www.opencalais.com/
[14] http://www.ai.sri.com/software/Chimaera
[15] https://stanfordnlp.github.io/CoreNLP/
[16] https://uima.apache.org/
[17] General Architecture for Text Engineering : https://gate.ac.uk/

The GATE family of tools has grown over the years to include a desktop client for developers, a workflow-based web application, a Java library, an architecture and a process. GATE is:

- an IDE, GATE Developer: an integrated development environment for language processing components bundled with a very widely used Information Extraction system and a comprehensive set of other plugins : see Fig. 3 the annotation interface of the software
- a cloud computing solution for hosted large-scale text processing, GATE Cloud (https://cloud.gate.ac.uk/).
- a web app, GATE Teamware: a collaborative annotation environment for factory-style semantic annotation projects built around a workflow engine and a heavily-optimized backend service infrastructure.
- a multi-paradigm search repository, GATE Mímir, which can be used to index and search over text, annotations, semantic schemas (ontologies), and semantic meta-data (instance data). It allows queries that arbitrarily mix full-text, structural, linguistic and semantic queries and that can scale to terabytes of text.
- a framework, GATE Embedded: an object library optimized for inclusion in diverse applications giving access to all the services used by GATE Developer and more.
- an architecture: a high-level organizational picture of how language processing software composition.

GATE as an architecture suggests that the elements of software systems that process natural language can usefully be broken down into various types of component, known as resources. Components are reusable software chunks with well-defined interfaces, and are a popular architectural form, used in Sun's Java Beans and Microsoft's .Net, for example. GATE components are specialized types of Java Bean, and come in three flavors:

- LanguageResources (LRs) represent entities such as lexicons, corpora or ontologies;
- ProcessingResources (PRs) represent entities that are primarily algorithmic, such as parsers, generators or ngram modellers;
- VisualResources (VRs) represent visualization and editing components that participate in GUIs.

These definitions can be blurred in practice as necessary.

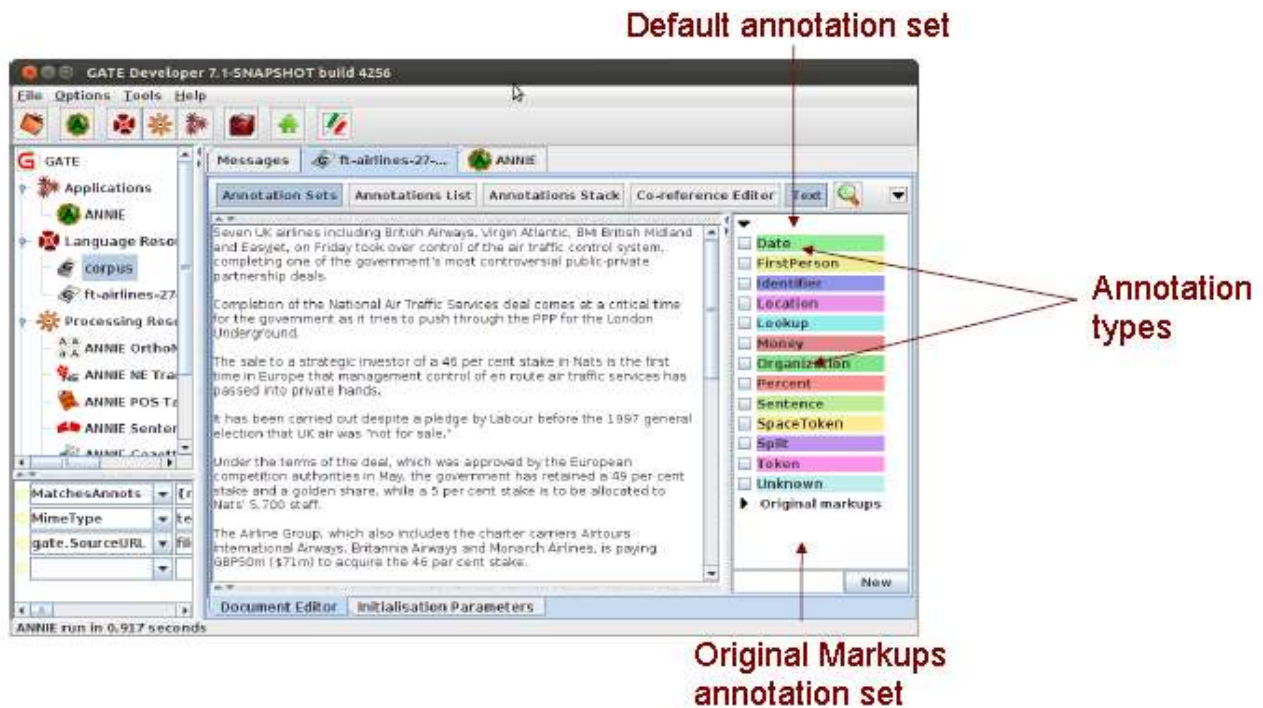

Fig. 3: GATE annotation interface (from GATE Courses, 2017)

## 2.5 Semantic annotation and standardization

We provide here a rough description of the ISO 24617-6 standard, since this part specifically deals with semantic annotation.

The concept of semantic annotation has been the subject of standardization work, developed by the ISO TC 37 SC4 committee. ISO 24617[18] consists of several parts, under the general title Language Resource Management — Semantic Annotation Framework (SemAF):

    — Part 1: Time and events (SemAF-Time, ISOTimeML)
    — Part 2: Dialogue acts (SemAF-Dacts)
    — Part 4: Semantic roles (SemAF-SR)
    — Part 5: Discourse structures (SemAF-DS) [Technical Specification]
    — Part 6: Principles of semantic annotation (SemAF Principles)
    — Part 7: Spatial information (ISOspace)
    — Part 8: Semantic relations in discourse (SemAF DR-core)
    — Part 9: Reference (ISOref)

ISO 24617-6: 2016 specifies the approach to semantic annotation characterizing the ISO Semantic annotation framework (SemAF). It outlines the SemAF strategy for developing separate annotation schemes for certain classes of semantic phenomena, aiming in the long term to combine these into a single, coherent scheme for semantic annotation with wide coverage. In particular, it sets out the notions of both an abstract and a concrete syntax for semantic annotations, mirroring the distinction between annotations and representations that is made in the ISO Linguistic Annotation Framework. It describes the role of these notions in relation to the specification of a metamodel and a semantic interpretation of annotations, with a view to defining a well-founded annotation scheme.

ISO 24617-6:2016 also provides guidelines for dealing with two issues regarding the annotation schemes defined in SemAF-parts: a) conceptual and terminological inconsistencies that may arise due to overlaps between annotation schemes and b) the treatment of semantic phenomena that cut across SemAF-parts, such as negation, modality and quantification. Instances of both issues are identified, and in some cases, direction is given as to how they may be tackled.

Among the terms defined in this part:
- **semantic annotation :** annotation which contains information about the meaning of a segment or region of primary data
- **primary data :** electronic representation of text or communicative behavior
A note follows the definition, this note is important to define the application domain of an annotation: "Note 1 to entry: ISO 24612 defines primary data as the 'electronic representation of language data'. This definition is unsatisfactory for this part of ISO 24617 as semantic annotation may relate to non-verbal or multimodal data, such as stretches of spoken dialogue with accompanying gestures and facial expressions, and even gestures and/or facial expressions without any accompanying speech." (ISO 24617-6, 2016)

In another domain, the annotation of standards has been the subject of a document from OGC (Open Geospatial Consortium Inc.[19]), under Reference OGC 08-167r1, in 2009, entitled "Semantic annotations in OGC standards". Even though this document is not considered by OGC as an official position of the OGC membership, it shows the importance of this approach to attach meaningful descriptions to the standards, thus increasing the usefulness of geospatial information.

In the next sections, we describe the ontologies developed by the authors in the domain of standardization and industrial information, we also outline the different ontologies, dictionaries, databases available, then their application to the semantic annotation of the ISO 15531 standard, the methodology and the results obtained so far.

---

## 3- Ontologies in the world of standardization

In this paragraph, we outline several work where ontologies and standardization are closely related, but also other work connected to the management of information and knowledge leading to the development of ontologies that could be very useful for representing industrial information.

### 3-1 Standards and standardization ontology: isto.owl

At the international level, standardization and standards are concepts well-defined through several guides, produced by both ISO and IEC, which are, for the most commonly used:
- ISO/IEC Guide, Standardization and related activities – General vocabulary (8[th] ed, 2004); (ISO/IEC Guide 2, 2004)
- ISO/IEC Directives, Parts 1 and 2 and the Consolidated ISO supplement: ISO/IEC Directives, Part 1, 13[th] Ed, 2017, ISO/IEC Directives, Part 2, 7[th] Ed, 2016, ISO Part 1 - Consolidated ISO supplement, 8[th] ed, 2017.
Those documents, particularly the ISO/IEC Guide, provide very precise definitions for all the terms related to standards and to the standardization process, among which the terms : (ISO/IEC Guide 2:2004)

- **standard**: document established by consensus and approved by a recognized body, that provides, for common and repeated use, rules, guidelines or characteristics for activities or their results, aimed at the achievement of the optimum degree of order in a given context;
- **standardization**: refers to the activity of establishing, with regard to actual or potential problems, provisions for common and repeated use, aimed at the achievement of the optimum degree of order in a given context.

Terms defined in the ISO/IEC Guide 2 are organized into categories, among which: standardization, aims of standardization, normative documents (for the different types of standards), bodies responsible for standards and regulations, types of standards. According to the ISO/IEC Guide 2, terms expressing more specific concepts may generally be constructed by a combination of terms representing more general concepts. The latter terms thus form "building blocks", and the selection of terms and the construction of definitions within this Guide have been based on those cases where equivalent English, French and Russian (official languages of ISO/IEC documents) combined terms contain the same "building blocks".

We have developed an ontology of standards and standardization (isto.owl): to date, the ontology consists of 125 classes, 44 object properties (used for specifying axioms) and 4 datatype properties (Cutting-Decelle et al., 2014). The development of the ontology is being done on a manual basis, through the analysis of ISO Directives and based on the authors' skill and experience as either members of SDOs or experts in standardization. This ontology is intended to be used as a common vocabulary applicable to the representation of whatever kind of standardization document, or activity, in order to make the information contained widely shareable and interoperable. The criteria used for structuring the ontology have been to keep as far as possible the duality of the standardization concept: both a method (the standardization process) and a result (the standard being developed). Given the complexity of the standardization world, and in order be as complete as possible with respect to the regulatory documents, the writing process of the set of axioms is still in progress.

An excerpt of the ontology, developed by the authors using the Protégé[20] software tool is represented Fig. 4 below:

---

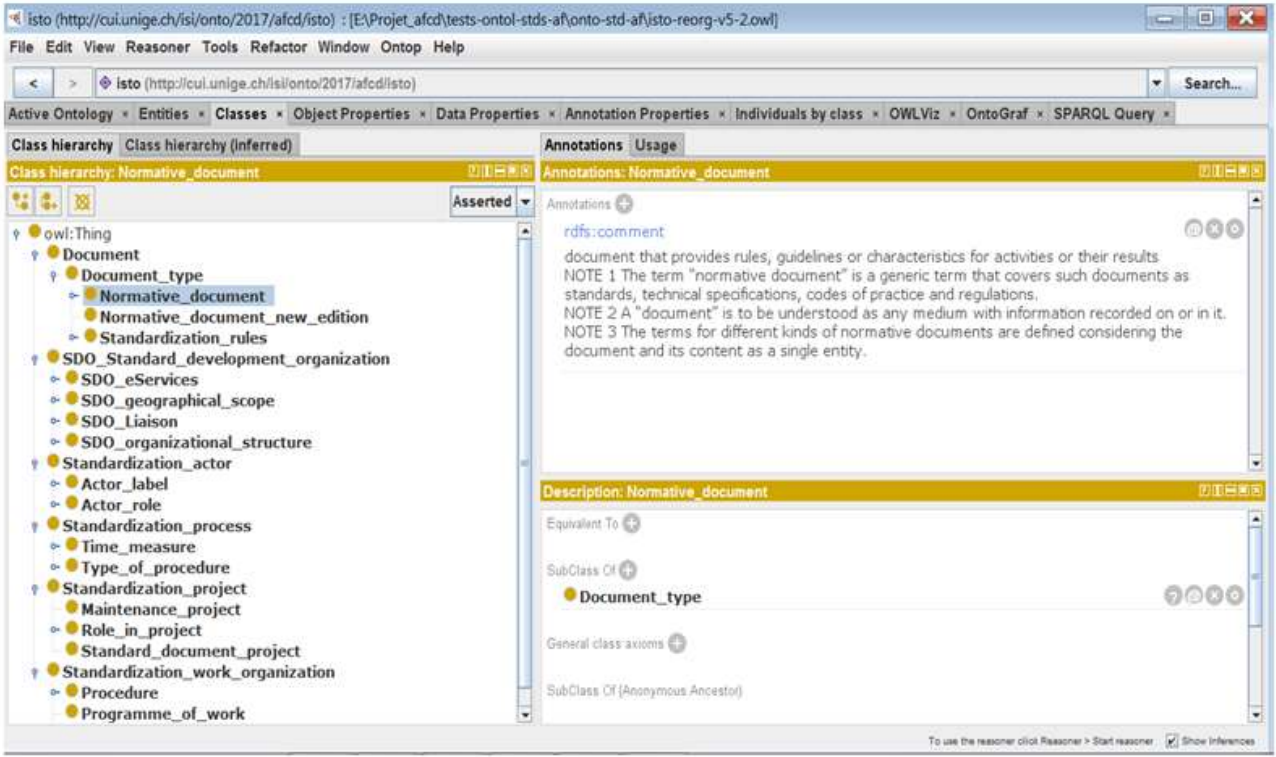

Fig. 4 : Standardization and standards ontology represented using Protégé 5.2 (excerpt) (isto.owl)

## 3-2 Reference and domain ontologies

Several initiatives leading to the development of ontologies, or technical dictionaries have been identified, the list we mention here cannot be exhaustive since an important part of the work done only appears in PhD reports, or in Master Thesis, not necessarily widely disseminated once the work is completed.

We must note, however, that the scope of most of the ontologies and vocabularies mentioned in this paper is wider than the mere information needed by the semantic annotation of the ISO 15531 MANDATE standard. Since this standard is linked to many other standardization work, and, more generally, linked to several knowledge domains, it seemed to us important to enlarge our approach to other, although closely related, technical domains.

### 3-2-1 Technical or domain ontologies

Most of the ontologies cited here come from the LOV (Linked Open Vocabulary) repository[21], they can be downloaded in .owl (or sometimes, .rdfs, .ttl or .n3),

- **Product data sheet ontology**, developed by NIST (US) (NISTIR 8035, 2014): Data exchange transactions for the engineering, procurement, and operation of process equipment depend heavily on product data sheets. Product data sheets are traditional paper forms that serve as the main method of communicating detailed equipment specifications. There are many standards for the layout and use of product data sheets, but there are still problems with the data exchanges that use them. First, most transactions are still done with the exchange of document images, e.g., scanned product data sheets, which have to be interpreted by humans and transcribed for use in a software tool. The ontology is a formal machine-readable terminology that is based on industry standard glossaries of terms and definitions. The ontology specifically includes the concepts used in Centrifugal Pump, Pressure Transmitter, and Valve data sheets, and depicts the product data sheet concepts and their relationships in diagrams in the Unified Modeling Language (UML), with a detailed description in text. The text includes the relationships of the ontology elements to industry standards for product data sheet information.
- **provoc**: (Product Vocabulary) is a vocabulary developed by INRIA (F) that can be used to represent information and manipulate them through the Web. This ontology reflects: 1) The basic hierarchy of a company: Group (Company), Divisions of a Group, Brand names attached to a Division or a Group, and 2) The production of a company: products, ranges of products (attached to a Brand), the composition of a product, packages of products..

---

[21] http://lov.okfn.org/dataset/lov/

Available at the URI : http://ns.inria.fr/provoc

- o **RAMI** : A vocabulary which represents the Reference Architecture Model for Industry 4.0 (RAMI), including the concept of the Administration Shell I4.0 Component
  Available at the URI : https://w3id.org/i40/rami/
- o **sto** : A vocabulary to describe the relation between standards in use on the Industry 4.0, and related concepts
  Available at the URI : https://w3id.org/i40/sto#
- o **smg** : A Profile of the IEC Common Information Model for Smart Grids, developed by the Cerise-SG project.
  Available at the URI : http://ns.cerise-project.nl/energy/def/cim-smartgrid
- o **aml** : A vocabulary to represent the AutomationML Standard - IEC 62714
  Available at the URI : https://w3id.org/i40/aml
- o **saref** : The Smart Appliances REFerence (SAREF) ontology is a shared model of consensus that facilitates the matching of existing assets (standards/protocols/data models/etc.) in the smart appliances domain. The SAREF ontology provides building blocks that allow separation and recombination of different parts of the ontology depending on specific needs
  Available at the URI : https://w3id.org/saref
- o **seas** : This vocabulary is version v0.1 of the ITEA2 Smart Energy Aware Systems project vocabulary. It enables the description of electricity measurements of a site using the Data Cube W3C vocabulary, (with several related ontologies : seas-sys, seasb, seasd, seast, seasto, seas-stats, seas-qudt, seas-op)
  Available at the URI : https://w3id.org/seas/
- o **ssn** : This ontology describes sensors, actuators and observations, and related concepts. It does not describe domain concepts, time, locations, etc. these are intended to be included from other ontologies via OWL imports.
  Available at the URI : http://www.w3.org/ns/ssn/
- o **ozone ontology** : The OZONE ontology and toolkit represent a synthesis of extensive prior work in developing constraint-based scheduling models for a range of applications in manufacturing, space and transportation logistics (https://www.cs.cmu.edu/afs/cs/project/ozone/www/AAAI_Symp_On_Ontol_97/abstract.html).
- o **proton ontology** : PROTON (PROTo ONtology) was developed in the SEKT[22] project as a lightweight upper-level ontology, serving as a modeling basis for a number of tasks in different domains. PROTON is meant to serve as a seed for ontology generation (new ontologies constructed by extending PROTON; it can be used for automatic entity recognition and more generally Information Extraction (IE) from text, for the sake of semantic annotation (metadata generation). (https://ontotext.com/products/proton/

### 3-2-2 Ontologies as standards

Several research work or industrial projects have contributed to the development of reference ontologies which have been standardized, among which:

- ISO/PRF 20534 (ISO/PRF 20534, 2018): developed by ISO TC 184 SC4, Industrial automation systems and integration

- Formal semantic models for the configuration of global production networks: ISO 20534 specifies a formal logic based concept specialisation approach to support the development of manufacturing reference models that can underpin the necessary business specific knowledge models that are needed to support the configuration of global production networks. The following are within the scope of ISO 20534: production networks for discrete product manufacture, formal semantics for the configuration of global production networks, system level formal semantics, designed system formal semantics, manufacturing business system formal semantics, global production systems network formal semantics (ref). ISO 20534 is developed within the framework of the EU FoF Project FP7-2013-NMP-ICT-FOF FLEXINET[23].

- ISO 18629 PSL: Industrial automation systems and integration (ISO TC 184 SC4) – Process Specification Language: International standard for providing semantics to the computer-interpretable exchange of information related to manufacturing and other discrete processes. Taken together, all the parts contained in PSL provide a language for describing processes throughout the entire production within the same industrial company or across several industrial sectors or companies, independently from any particular representation model. The nature of this language makes it suitable for sharing process information during all the stages of production (ISO 18629-1, 2004).

- ISO/TS 15926-12:2018[24] (ISO TC 184 SC4) specifies an ontology for the integration of industrial data throughout its life-cycle. The ontology is represented in OWL (W3C).

---

[22] http://www.sekt-project.com/
[23] http://www.flexinet-fof.eu/pages/flexhome.html
[24] https://www.iso.org/standard/70695.html

- ISO/IEC 21838[25] : series of standards developed by ISO/IEC JTC1/SC32, whose parts are currently under development: ISO/IEC CD 21838-1 (Information technology, Top-level ontologies, Part 1: Requirements) and ISO/IEC CD 21838-2 (Information technology, Top-level ontologies, Part 2: Basic Formal Ontology (BFO)).

In the domain of geographic / geomatics information, the ISO TC 211[26] (Geographic information/Geomatics) committee has developed several ontologies, known as ISO 19150: ISO/TS 19150-1:2012 (reviewed and confirmed in 2016) defines the framework for semantic interoperability of geographic information. This framework defines a high level model of the components required to handle semantics in the ISO geographic information standards with the use of ontologies. The other parts of this standard, either already developed or foreseen are : Part 2: Rules for developing ontologies in the Web Ontology Language (OWL), Part 3: Semantic operators, Part 4: Service ontology, Part 5: Domain ontology registry, Part 6: Service ontology registry.

In another domain, semantic approaches and ontologies are also used for information exchange applied to financial services (ISO/TC 68/SC9).

### 3-3 Dictionaries, glossaries, vocabularies and databases

- **Electropedia** (http://www.electropedia.org/) : The World's Online Electrotechnical Vocabulary. Electropedia is produced by the IEC, the world's leading organization that prepares and publishes International Standards for all electrical, electronic and related technologies – collectively known as "electrotechnology". Electropedia (also known as the "IEV Online") contains all the terms and definitions in the International Electrotechnical Vocabulary or IEV which is published also as a set of publications in the IEC 60050 series that can be ordered separately from the IEC webstore. Electropedia is the world's most comprehensive online terminology database on "electrotechnology", containing more than 22 000 terms and definitions in English and French organized by subject area, with equivalent terms in various other languages: Arabic, Chinese, Czech, Finnish, German, Italian, Japanese, Korean, Norwegian (Bokmål and Nynorsk), Polish, Portuguese, Russian, Serbian, Slovenian, Spanish and Swedish (coverage varies by subject area). The world's experts in electro-technical terminology work to produce Electropedia under the responsibility of IEC Technical Committee 1 (Terminology), one of the 203 IEC Technical Committees and Subcommittees.

- **Online Browsing Platform (OBP)** (https://www.iso.org/obp/ui#home): provides access to the most up to date content in ISO standards, graphical symbols, codes or terms and definitions. Preview content before you buy, search within documents and easily navigate between standards, according to the following criteria: standards, collections, publications, graphical symbols, terms and definitions, country codes.

- **Common Data Dictionary** (IEC CDD 61360) is a database maintained by IEC SC3D: "Product properties and classes and their identification". It is a common repository of concepts for all electrotechnical domains based on the methodology and the information model of IEC 61360 series, and provides : an unambiguous identification of classes and properties, and their relations; a commonly accepted terminology and definitions based on accepted sources such as IEC International Standards, other International Standards, industry standards, or public authorities; hierarchies of concepts enabling users to appropriately characterize their products and services; relevant conditions and constraints if necessary on possible values of characteristics; technical representation of concepts including units and data types and their identification. See : https://cdd.iec.ch/cdd/iec61360/iec61360.nsf

- **SC5 glossary**: the document results from the work of the ISO TC 184 SC5 study group, in 2008. The intention of this document is to guide SC5 working groups and other interested parties in their standardization work by providing them with an overview of definitions already available in SC5 standards (SC5 glossary, 2008) (Michel, 2005), (Kosanke, Martin, 2008).

---

## 3-4 Joint standardization working groups and non-profit organizations

- Joint standardization working groups are currently working at the international level (both ISO and IEC) to try to harmonize terms and definitions used in the domain of smart manufacturing, based on several national contributions in terms of vocabulary, among which RAMI (IEC PAS 63088, 2017), Big Picture (ISO/TR 23087, 2018) and other contributions from Japan (Proposal of Smart Manufacturing reference model) and China. This work is important since the terms mentioned in the different documents are more or less the same, but the definitions can be very different, and sometimes the approaches described in the documents not really compatible, nor interoperable !

- Another valuable project is carried out by IOF[27] : the primary goal of the IOF (Industrial Ontologies Foundry) is to create a suite of *open* and principles-based ontologies, from which other domain dependent or application ontologies can be derived in a modular fashion, remaining 'generic' (i.e., non-proprietary, non-implementation specific) so they can be reused in any number of industrial domains or manufacturing specializations. The other goals of IOF are : Providing principles and best practices by which quality ontologies can be developed that will support interoperability for industrial domains, Instituting a governance mechanism to maintain and promulgate the goals and principles, Providing an organizational framework and governance processes that ensure conformance to principles and best practices for development, sharing, maintenance, evolution, and documentation of IOF ontologies.

The following notional diagram (Fig. 5) suggests how the suite of proposed IOF ontologies and their progeny may evolve. The expectation is that the application and/or bridging ontologies would be private or perhaps licensed.

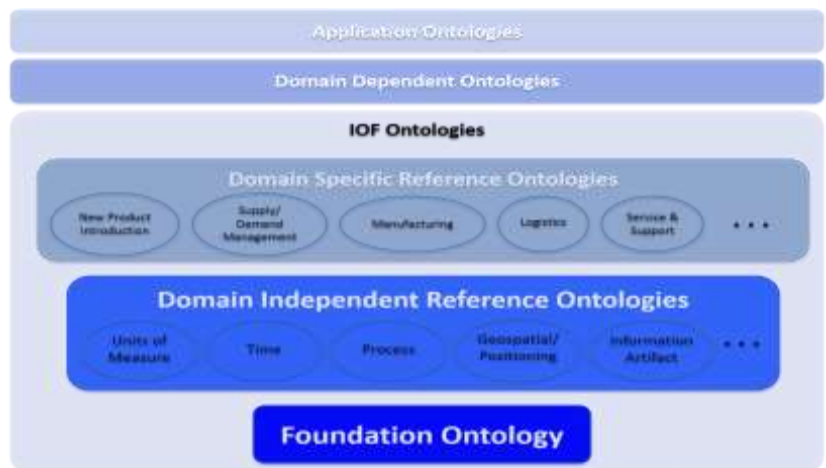

Fig. 5: different levels of IOF ontologies (ref IOF)

## 3-5 Specificity of the documents generated by the standardization process

An important point needs to be considered at this stage: standards, either at the national level, European and international levels are not free documents. They are drafted either by NSBs, or SDOs, then sold by the same organizations. In addition, normative documents are the subject of systematic reviews (ISO Guidance, 2017): every International Standard published by ISO alone, or jointly with the IEC, is subject to Systematic Review (SR) in order to determine whether it should be confirmed, revised/amended, converted to another form of deliverable, or withdrawn (ISO Supplement to the ISO/IEC Directives, Subclauses 2.9.1 and 2.9.2.). The committee can decide to launch the SR whenever necessary, or it is automatically launched 5 years after the publication or confirmation of an International Standard (IS).

Note: SR ballots for IS are automatically launched after 5 years and for Technical specifications (TS) and Publicly Available Specifications (PAS) after 3 years. SR ballots for other deliverables are only sent on request.

For all these reasons, it is not very easy to work on normative documents, since it is, first of all, necessary to buy them, then to be sure to work on the latest version.

---

As a matter of introduction to the next sections, it is interesting at this stage to mention the research work done by A. Fraga, M. Vegetti and H. Leone. In their work (Fraga et al., 2016), (Fraga et al., 2016), (Fraga et al. 2017), they have built a three-levels architecture, based on ontologies, whose aim is to improve the interoperability between ISO TC 184 SC4 standards. For their analysis, they mainly focused on the following standards: ISO 10303, ISO 13584 and ISO 15531, and particularly on the definitions of the terms of "product", "process" and "resource". In one of their paper, they propose a semi-automated ontology generation process from product data standards based on UIMA (Unstructured Information Management Architecture). One of the problem they face is of course the alignment of the ontologies obtained from the extraction process: whence the justification of the need for a common dictionary of the terms handled in the documents !

## 4- Extraction of information from a specific normative corpus based on a semantic annotation process: methodology and application to ISO 15531 MANDATE

This section starts with the description of the main features of the standard, its scope and its structure. Given the domain covered, the standard makes use of terms, concepts, either specific to the domain of manufacturing management, or coming from industry, or else more generic. Those considerations have led us to create separate ontologies (if they did not exist previously) or to re-use existing stuff. The third part of this section describes the annotation process of the standard, based on the use of the GATE text engineering software, the different stages of the methodology and a first application to the standard.

### 4-1 Overview of the ISO 15531 MANDATE standard

A manufacturing management system manages the flow of materials and products through the whole production chain, from suppliers, through manufacturers, assemblers, to distributors, and sometimes customers. The relations among those partners may be identified and structured in an electronic form with a view to facilitate electronic exchanges. Then, information handled during these exchanges have to be identified, modeled, and represented in such a way that they may be shared by a maximum of partners. From this analysis, three main categories of data related to manufacturing management can be distinguished (ISO 15531-1, 2004), (Cutting-Decelle et al., 2009) :

- information related to the management of time;
- information related to the management of the resources used during the manufacturing processes;
- information related to the management of manufacturing flows.

ISO 15531 MANDATE (MANufacturing management DATa Exchange) is an International Standard for the computer-interpretable representation and exchange of industrial manufacturing management data. The nature of the description makes it suitable not only for neutral file exchange, but also as a basis for implementing and sharing manufacturing management databases and archiving. The standard is focused on discrete manufacturing, but not limited to it. The purpose is to facilitate the integration between the numerous industrial applications by means of common, standardized software tools able to represent these three sets of data.

The standard is organized as a series of parts, each published separately. The parts belong to the following series: (IS: International Standard)

- **Manufacturing resources usage management data (3x series) :**
  – **ISO IS 15531-31**: Resource Information Model: Basic Concepts. This part is an introduction to the ISO 15531-3x series of part of ISO 15531. It describes the universe of discourse of this standard as well as the resources information model. It provides the main principles used in this series of parts of ISO 15531 and presents the fundamental principles used for the conceptual model of resource usage management data.
  – **ISO IS 15531-32**: Conceptual Model for Resources Usage Management Data : The conceptual information model for resources usage management data is structured into six logical modules, which are: resource hierarchy (generic, specific, individual resource), resource characteristics (set of information about a resource), resource administration (administrative information), resource status (availability or not of the resource), resource view (specific aggregation of resources), resource representation (physical values), resource configuration. A resource is the basic element for resource management. Each further detailed description, classification or configuration of resources relates to resource. A resource can be generic, specific, or individual and may in turn be made of a number of other resources. Each resource has characteristics and can also be considered from different viewpoints. It is important to notice that a resource is not a priori related to any given activity. It exists and may be managed before any appointment to any activity. That is typically the case for human resources.

- **Manufacturing flow management data (4x series) :**
  – **ISO IS 15531-42**: Time Model. The time schema provides the definition of concepts related to the time representation, needed by software applications mainly dealing with scheduling and manufacturing management operations. It enables multiple representations of time domains, intervals of time, points in time, and time units.
  – **ISO IS 15531-43**: Data Model for Manufacturing Flow Management. This part addresses the modeling of data for the management of manufacturing flows as well as flow controls in a shop floor or in a factory. This manufacturing flow model is provided in the context of various processes that run simultaneously and/or sequentially, providing one or more products and/or components and involving numerous resources. This part provides a way to model the data needed to manage the multiple complex flows that have to be taken into account between the different manufacturing processes in a factory. That includes products, components, or raw material flows as well as services flows, such as information flows.
  – **ISO IS 15531-44**: Shop Floor data for Manufacturing Management (ISO 15531-44) : This part addresses the modelling of the data collected from data acquisition systems at control level to be stored at the manufacturing management level and processed further at this level for any management purpose (Cutting et al., 2012). The following features belong to the scope of the standard: quantitative or qualitative data collected from data acquisition systems at the control or management level to be stored at the management level and used later on to manage manufacturing, time stamping and time measurement provided from data acquisition systems for control and management data. According to the fact that the model will have to be as generic as possible and easy to specialize, the entities described are themselves very generic. Their specialization, when needed, is obtained through the use of PLIB libraries (ISO 13584-1 [10] and ISO 13584-24). In that case the specialization process is roughly described in the standard.

ISO 15531 MANDATE Part 1 provides a general overview, specifying the functions of the various series of parts of the standard and the relationships among them. It also specifies the relations between the standard and other related standards. All the MANDATE parts are written using the EXPRESS language[28].

**4-2 Identification of the ontologies and databases of interest**

The normative corpus used for this study is made up of 6 documents, wich are the 6 parts of the standard :
- Part 1 : General overview (32 pages);
- Part 31 : Resource information model (29 pages);
- Part 32 : Conceptual model for resources usage management data (37 pages);
- Part 42 : Time model (43 pages);
- Part 43 : Data model for flow monitoring and manufacturing data exchange (25 pages);
- Part 44 : Information modelling for shop floor data acquisition (31 pages).

A careful reading of the 6 documents shows different domains of knowledge. Given the scope and coverage of the whole ISO 15531 standard, we have identified 4 domains in our study, which are :
- generic information about standards;
- information provided by the ISO and IEC databases;
- technical knowledge external to the domain of the standard;
- knowledge specific to the technical domain covered by the standard.

- **Generic information about standards** : relates to terms specific to the vocabulary of standards and standardization, as dealt with by the isto ontology (see section 3-1), such as "International standard", "scope", "part" and more generally all the terms defining the structure and the normative environment of the document. This information has been represented in the isto ontology, see Fig. 4.

- **Information provided by ISO, IEC, CEN databases**: such as the numbering and naming of the normative documents: "ISO 15531", "ISO 10303", "Industrial automation system and integration – Product data representation and exchange – Part 1: Overview and fundamental principles". This information is managed by the standardization bodies and available on their web sites (see Fig. 8).

Those first two categories are relatively easy to identify, since they correspond to what can be called an "objective" reading of the document.

The identification of knowledge used in the two other categories below is not so straightforward since, as we will see it in the next sections, terms can be borrowed from a given domain (or corpus), then used in another, but with (more or less) small changes in their definitions ! Examples of those ambiguities will be provided in the next section.

- **Technical knowledge external to the domain of the standard**: we put in this category terms defined either in other standards developed by SC4, or else by ISO/TC 184 (with a scope wider than SC4's), or possibly by other TCs, other standardization bodies, even by non-profit organizations for some of them. The difficulty is here to be able to identify those terms, then to identify the document (or documents!) where they are defined, but also to identify possible synonymies, and eventual mistakes. This work is leading to the development of one or several ontologies, with serious problems in the definitions of the concepts throughout the big number of documents to process, whence a need for a strong alignment among the ontologies. Working on ontology based approaches makes it possible, through the software tools used, to check the consistency of the ontologies as they are built – since the number of classes, axioms, relations can become very big!

- **Knowledge specific to the technical domain covered by the standard**: the scope of ISO 15531 being manufacturing management, this category includes all the terms of vocabulary dealing with production management: product (as something to be manufactured), manufacturing process, resources, time, all data and information related to the production process. Scope of ISO 15531 is: specification of the characteristics for a representation of manufacturing management information over the entire industrial process. It provides the necessary mechanisms and definitions to enable manufacturing management data to be shared and exchanged within the factory, with other plants or companies. The standard is mainly focused on discrete manufacturing but not limited to it (ISO 15531-1, 2004).

Today the ontologies mentioned are being developed by hand, since the objective of the study is to set up a proof of concept enabling a validation of the proposed methodology.

### 4-2-1 Ontology of the ISO 15531 MANDATE standard: main features and issues
The ontology of ISO 15531 (iso15531.owl) represents the different classes defined in the 6 parts of the standard. An excerpt of the ontology developed by the authors is represented Fig. 6 below:

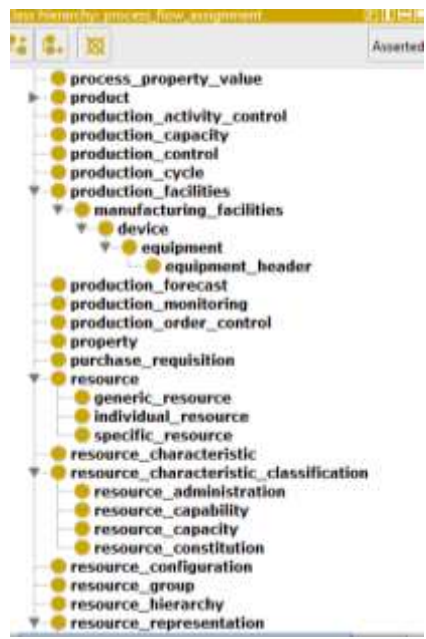

Fig. 6: excerpt of the ISO_15531 MANDATE ontology (Protégé 5.2) (iso15531.owl)

The structuring of the classes (sub-classes) is provided by the EXPRESS definition of a "SUBTYPE OF" clause in the standard.

### 4-2-2 Other technical or reference ontologies
This work is just starting, since it needs lots of references to other ontologies (related to other SC4's, or other TC standards). It is also the most difficult part of the work, with need for alignment of the different ontologies in relation.

For the work done here (tech.owl) by the authors, the main classes out of the scope of ISO 15531 are represented Fig. 7 below. For all the classes identified in the ontology, a reference is mentioned to the part/standard they come from.

We will describe in the next section some of the problems we are meeting with this referencing.

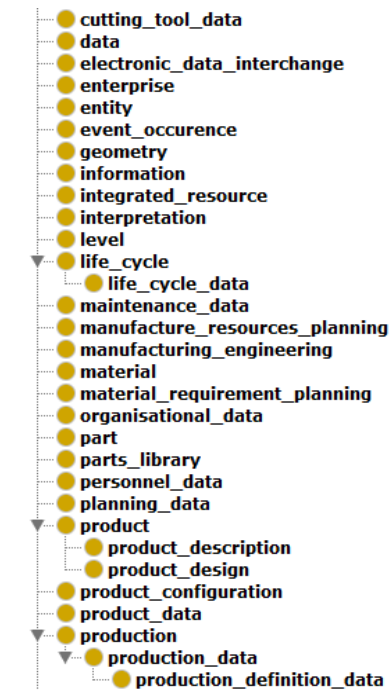

Fig. 7 : technical ontology (excerpt) : main classes (Protégé 5.2) (tech.owl)

### 4-2-3 Dictionaries, databases

An example of database accessible on the web is provided on the Fig. 8 below: the Online Browsing Platform (OBP) developed by ISO enables the identification of some (public) information about standards, based on key-words. The user of the platform can select different categories (shown on the left panel of the web wite) and the results of the query (here on the term of "resource") are shown together with the references of the standards where the term appears.

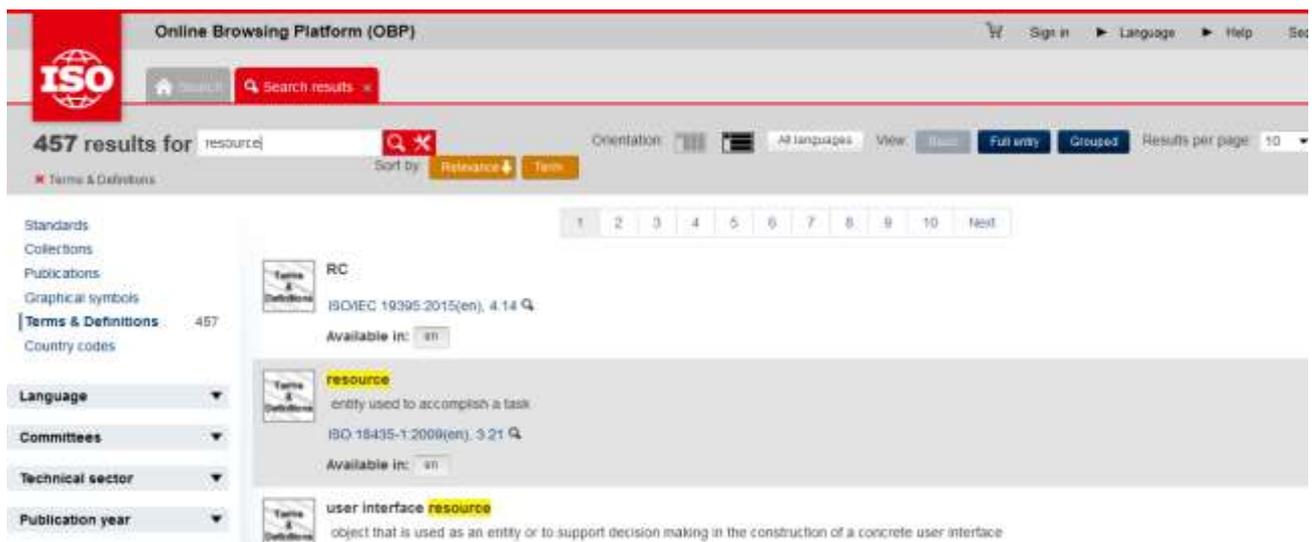

Fig. 8 : ISO Online Browsing Platform (OBP)[29]

The previous ontologies provide the basic vocabulary (and relations) used to extract technical information from the ISO 15531 parts. In the next section, we present the methodology, then some of the first results obtained so far.

---

[29] https://www.iso.org/obp/ui/#search

**4-3 Information extraction using GATE: methodology and application to the ISO 15531 parts**

For our work, we are using GATE Developer (v. 8.4.1), the GATE graphical user interface. Core concepts are[30] :
- o   the documents to be annotated,
- o   corpora comprising sets of documents, grouping documents for the purpose of running uniform processes across them,
- o   annotations that are created on documents,
- o   annotation types such as 'Name' or 'Date',
- o   annotation sets comprising groups of annotations,
- o   processing resources that manipulate and create annotations on documents, and
- o   applications, comprising sequences of processing resources, that can be applied to a document or corpus.

GATE includes a complete information extraction system free to use, called ANNIE (a Nearly-New Information Extraction System). GATE can load either documents, or sets of documents, or corpus (corpora). Documents, or corpora a considered as "Language resources" which will be the subject of the annotation process.
Those annotations can be defined "by hand", or provided by pre-defined lists, or else by ontologies. The document editor also provides a co-reference editor (for example, correspondence between "him" and "Paul" according to the context of the text).
GATE provides an API for modeling and manipulating ontologies and comes with two plugins that provide implementations for the API and several tools for editing ontologies and using ontologies for document annotation. Ontologies in GATE are classified as language resources. In order to create an ontology language resource, the user must first load a plugin containing an ontology implementation (here Ontology_OWLIM2).

The annotation process based on ontologies is done in several steps:

- creation of the set of documents, or the corpus and population with the parts of the standard,
- integration of the parts as language resources (LR),
- if the ontology plugin is loaded, integration of the ontologies as LR (if they are saved in RDF/XML),
- annotation of the part according to the ontology, using OAT (ontology annotation tool, plugin to manually annotate a text) : the set of annotations is related to the ontology displayed on the right of the screen. The choice of the colors for the annotations may be modified by the user.

It is possible to develop the ontology directly with GATE, used as ontology editor. For our work, ontologies have been previously written with Protégé 5.2.

Fig. 9 and 10 below show 2 examples of annotation using GATE, on the same "page" of the text, according to the ontology, the concepts annotated are not the same.
- Fig. 9: ISO 15531-44 with the ontology iso15531.owl
- Fig. 10: ISO 15531-44 with the ontology tech.owl

---

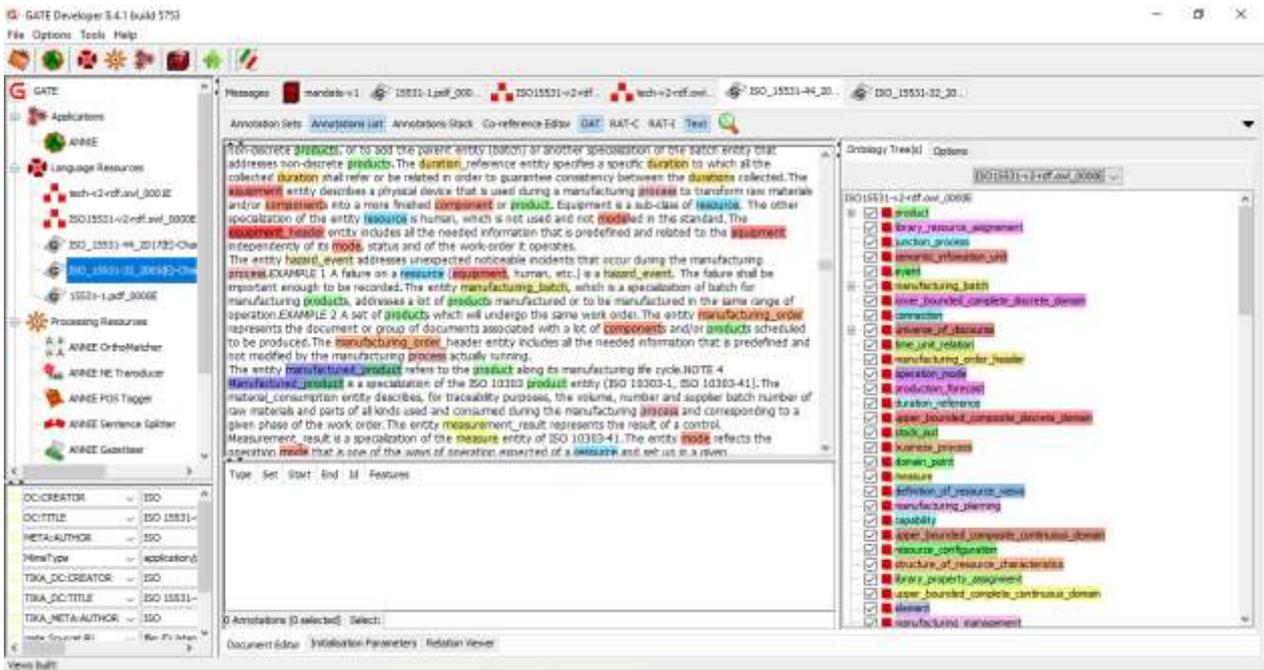

Fig. 9 : Annotation of ISO 15531-44 with the iso15531.owl ontology

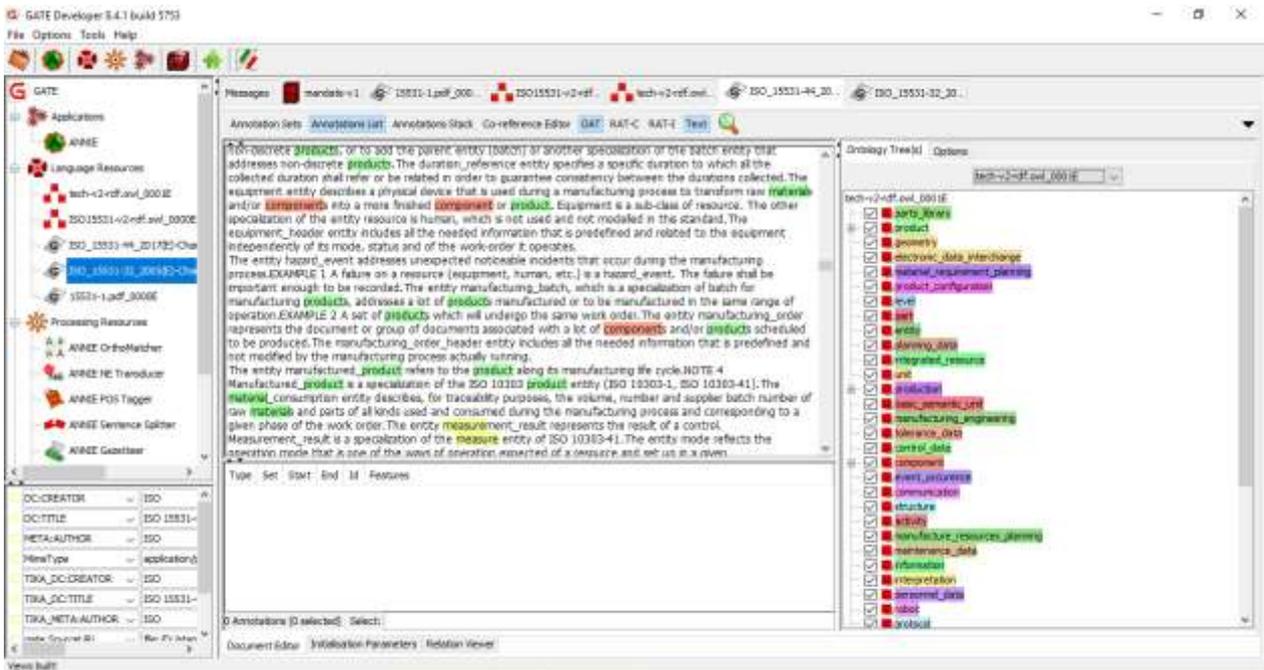

Fig. 10: Annotation of ISO 15531-44 with the tech.owl ontology

## 5- Discussion, analysis of the first results

### 5-1 Discussion, first comments

In order to validate the approach, the methodology proposed for the semantic annotation of normative documents has been applied to a standard well known by the authors of this paper, as contributors to the drafting of the parts. The objective was, through the elaboration of a POC (Proof of concept) to check the feasibility of the approach, but also to identify problems, potential difficulties and/or impossibilities.

For this project, the necessary ontologies have been developed by hand (work still on-going) and progressively completed / checked (enriched) during the reading process of the Parts. With a software such as GATE, it is possible to automatically enrich the ontology used as reference for the annotation; it is also possible, through the use of specific plugins provided by the tool and rules (written in JAPE) to automate the annotation process – in that case, according to the complexity of the process and the number of rules, it may take a certain time (minutes) to compute.

It is also possible to create ontologies from IE (Information extraction) software tools: since IE an important part of NLP (Natural Language Processing), lots of work has been done in this domain and many tools (open source or not, web tools or not) are available, some of them being dedicated to more specialized vocabularies, such as biology, biomedicine, geography, archeology, others contributing to provide important recognition features to the domain of semantic web.

An important (even critical!) step of the process is the alignment of the ontologies obtained during, and at the end of the annotation process: when they are developed in a separate way, and independently, it is quite possible to obtain "incompatible" ontologies – in that case they are very difficult to align, and not very useful !
Whence the need to high level approaches, such as the approach proposed by IOF (Industrial Ontologies Foundry) (see section 3-4) in order to federate high level vocabularies applicable to e.g., industry.

However the annotation process shows a number of problems, for some of them closely related to the specificity of the documents, as normative documents, among those problems, intertextuality.
Intertextuality[31] is the shaping of a text's meaning by another text. It is the interconnection between similar or related works of literature that reflect and influence an audience's interpretation of the text. Intertextuality is a literary device that creates an 'interrelationship between texts' and generates related understanding in separate works. These references are made to influence the reader and add layers of depth to a text, based on the readers' prior knowledge and understanding. Intertextuality is a literary discourse strategy utilised by writers in novels, poetry, theatre and even in non-written texts (such as performances and digital media). Examples of intertextuality are an author's borrowing and transformation of a prior text, and a reader's referencing of one text in reading another. Intertextuality does not require citing or referencing punctuation (such as quotation marks) and is often mistaken for plagiarism. Intertextuality can be produced in texts using a variety of functions including allusion, quotation and referencing. However, intertextuality is not always intentional and can be utilised inadvertently.
Intertextuality is often met in normative documents, often intentional, often also without reference to the original text – it is not always easy to identify the "original" text / source!

Another difficulty: terms ambiguities, depending on the use of uppercase of lowercase: example "is" means the verb, "IS" means International Standard – not easy to identify since tokenisers usually separate verbs from other elements of a sentence. However, the objective of our work is not NLP, even though some of the problems we face look similar to problems met in NLP. Another example of ambiguity is of course the concept of "part" (as spare part, for example, as from ISO 13584 P-LIB) and "part" as a fascicle of a multi-part standard.

---

**5-2 Analysis of the first results**

The writing of an ontology, particularly the ontology of the MANDATE standard revealed surprises, among which some mistakes, either in the content (mutual referencing among the parts not necessarily consistent) or else in the presentation of the document itself: see the example, shown Fig. 11 below:

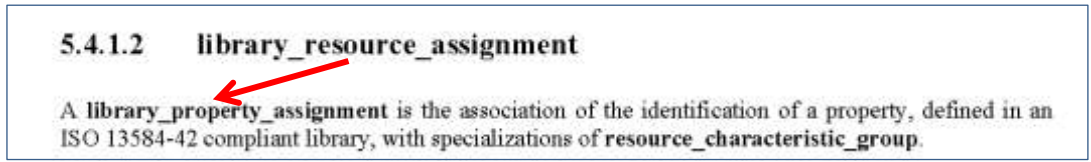

Fig. 11: Difference between the title and the content of a clause (p. 12) (ISO 15531-32:2005)

- Between ISO 15531-42 and ISO 15531-44: time related concepts are not always compatible.
- "basic_semantic_register": the definition in ISO 15531-1 is "adapted" from ISO TS 16668. This concept of "adaptation" appears in several clauses of nearly all the parts of the standard: but the "degree" of adaptation is not necessarily easy to interpret!
- "flow control" (ISO 15531-42) and "operation" (ISO 15531-43) are defined twice, in the part where they are used, and in ISO 15531-1: this is not necessary! – this double definition appears often enough throughout the parts, but not always with a perfect identity between the definitions !
- in the ISO 15531-42 part, several entities (related to time, time boundary, … ) are defined by multiple inheritance : not very well dealt with by ontology tools such as Protégé (mainly binary links).
- When entities are referred (for their definition) to standards outside the corpus, or to definitions outside the world of standardization (e.g. non-profit organizations), the definitions need to be carefully checked, and potential evolutions need to be followed in case of possible changes.

**Conclusions / perspectives of this work**

One of the main objectives of the study presented in this paper was to analyze the feasibility of the use of semantic approaches applied to normative documents.
The aim is to significantly influence and enhance information standards development in the future. This is needed to ensure cross standard consistency, a vital request to enable industry to use multiple sets of information standards, with the knowledge that all standards will interoperate and provide a consistent base for information sharing and exchange. This goal is something experts and standards users have been seeking for a long time.

We believe the approach developed here has such potential and could prove essential to perform the standardization work required for such big tasks as Industry 4.0, just to name one.

The consistency checking tool has already demonstrated results (see section 5-2), identifying inconsistencies across parts of a standard. The work on standardization led by the authors let us believe that this is a helpful tool, as, generally, for a multi-part standard, the different parts are not developed by the same persons, nor at the same time!

The methodology proposed here is widely applicable to TC 184 SC4 and TC 184 SC5 standards (our example, MANDATE is developed by ISO/TC 184 SC4/SC5), but also to other ISO TC 184 standards. It can also be envisaged to enlarge its use to other TC's (ISO and/or IEC, ITU …), depending on the existence (or the possibilities of development) of high level reference ontologies (Young et al., 2007), (Young et al., 2013), necessary to provide a common "hat" to the technical ontologies used in the annotation process of the normative documents.

In this article, we used 3 ontologies (and the database of standards related information provided by ISO), but according to the scope of the standard to be annotated, it can be necessary to include additional ontologies, and also additional databases, if needed.

Many interesting points can be derived from this work:

- an important potential way forward deriving from this work is the integration of automated annotation processes (with the enrichment of the ontologies used for the annotation) to software tools enabling a quick survey on standards, particularly when it comes to "big" standards, either in terms of number of pages and/or number of parts, according to criteria provided by the user.
This approach offers possibilities of tagging specific terms in the standards annotated, such as "smart manufacturing", or other terms well identified with a specific objective.

- Ontologies can be seen as specific "dictionaries" of specific domains. Our approach could offer the possibility for those ontologies to become official / legal documents, becoming a kind of a "reference dictionaries" of a given domain, thus providing any developer of a new standard within the domain productivity tools like a set of pre-defined, shared and applicable vocabularies. It could also help to improve interoperability among standards, and prevent from the development of "stand-alone" dictionaries of terms used by industry all over the world – but incompatible in between them.

- We have been here specifically working on standards developed by TC 184 and TC 184 SC4 which uses for drafting their documents a standardized language, EXPRESS (ISO 10303-11). But in any case, we strongly believe that formal semantic methods should play a significant part in the development of future industrial data standards.

More work is still required on this promising field.

*This work was conducted using the Protégé resource, which is supported by grant GM10331601 from the National Institute of General Medical Sciences of the United States National Institutes of Health.*

*The views expressed in this publication are those of authors and do not necessarily reflect the position or policies of the Standardization bodies mentioned nor the companies they work for.*

**Acknowledgements**: the ISO 15531 MANDATE corpus has been developed by experts of ISO TC 184 SC4 - SC5 JWG8, among which A.F. Cutting-Decelle (current convener), R.I. Young (deputy-convener), J.L. Barraud (expert); P. Lamboley is the current Chairman of ISO TC 184, A. Digeon was the previous Chairman and J.J. Michel the former JWG8 convener.

Intelligence and Data Mining, 0.1109/CIDM.2013.6597251